\documentclass[sigconf,nonacm]{acmart}
\usepackage{subcaption}
\usepackage{multirow}
\usepackage{placeins} 
\usepackage{float}
\usepackage{afterpage}
\usepackage{graphicx}
\usepackage{booktabs}
\usepackage{caption}

\AtBeginDocument{%
  }

\begin{document}

\title{Quantifying Gender Bias in Large Language Models: When ChatGPT Becomes a Hiring Manager}
\titlenote{This paper is based on Nina Gerszberg’s Master’s thesis submitted to the Massachusetts Institute of Technology (MIT) in 2024. The thesis is available at \url{https://dspace.mit.edu/handle/1721.1/156812}.}

\author{Nina Gerszberg}
\email{ninager01@alum.mit.edu}
\orcid{0009-0005-4452-5444}
\authornotemark[1]
\affiliation{%
  \institution{Massachusetts Institute of Technology}
  \city{Cambridge}
  \state{Massachusetts}
  \country{USA}
}

\author{Janka Hamori}
\email{hamorijanka12@gmail.com}
\affiliation{%
  \institution{Massachusetts Institute of Technology}
  \city{Cambridge}
  \state{Massachusetts}
  \country{USA}
}
  
\author{Andrew Lo}
\email{alo-admin@mit.edu}
\affiliation{%
  \institution{Massachusetts Institute of Technology}
  \city{Cambridge}
  \state{Massachusetts}
  \country{USA}
}

\renewcommand{\shortauthors}{Gerszberg et al.}

\begin{abstract}
  The growing prominence of large language models (LLMs) in daily life has heightened concerns that LLMs exhibit many of the same gender-related biases as their creators. In the context of hiring decisions, we quantify the degree to which LLMs perpetuate societal biases and investigate prompt engineering as a bias mitigation technique. Our findings suggest that for a given resumé, an LLM is more likely to hire a female candidate and perceive them as more qualified, but still recommends lower pay relative to male candidates. 

\end{abstract}

\maketitle

\section{Introduction}

Large language models (LLMs) have undergone substantial technical advancements, making them invaluable tools across various tasks and industries.  However, survey studies demonstrate that the cultivation of training data has resulted in many LLMs exhibiting a variety of biases. \cite{gallegos2023bias, sheng-etal-2021-societal, stanczak2021survey} 

Our work employs tools used by social scientists to examine biases in society and applies them to quantify gender-related biases in LLMs in the context of hiring decisions.
Bertrand and Mullainathan's ``Are Emily and Greg More Employable than Lakisha and Jamal" \cite{lakishaandJammal} — one of the most prominent societal bias studies — used names as a proxy for race, sending out identical resumés with different names to assess the impact of race on one's employability. We follow a very similar technique to quantify gender biases in LLMs. 

Our research quantifies the extent to which LLMs discriminate against job applicants on the basis of gender. A recent study \cite{greenhouse2023} found that 19\% of job applicants have changed their names primarily due to discrimination concerns. We utilize names as a gender proxy, presenting LLMs with identical resumés but different names to evaluate gender-based effects on the following three hiring-related questions: 1) How qualified is this candidate? 2) Would you hire them? and 3) What should their total compensation be? We also investigate the effectiveness of two prompt-based bias-mitigation techniques. First, we examine the results of instructing the LLM to articulate its reasoning behind its response; second, we ask the LLM to be as fair and equitable as possible and to consider the values of diversity, equity, and inclusion in its response. In addition to our experimental results, our paper proposes a concrete definition for bias as well as a methodology that can be easily replicated and modified to run additional experiments testing other forms of bias or other bias mitigation techniques.

The rise of large language models has revolutionized a number of industries, including human resources \cite{gustafson2023llms}. According to the Society of Human Resource Management, approximately one in four organizations use automation or artificial intelligence to throughout the hiring process, with many hiring managers using LLMs to grade resumes and decide which candidates to hire \cite{gan2024application}. Though 47\% of Americans believe that AI will outperform humans at treating job applicants in the same way \cite{pew2023}, it is unclear if AI systems are truly more equitable than humans. Furthermore, the ecological validity of bias resulting in LLM involvement in the hiring process is critical given laws such as New York City’s Local Law 144 requiring bias audits for Automated Employment Decision Tools (AEDT) \cite{nyc_aedt_notice}.  That being said, the specific hiring scenario provided here is secondary to the overarching methodology and insight it provides. Our results demonstrate that measuring an LLM's inherent, underlying biases is a nontrivial task requiring a number of probes from a variety of perspectives, including questions an LLM may not frequently encounter. These techniques revealed a number of biases in the way LLMs differentiate in their recommendations of job candidates based on gender. LLMs recommend paying women less than otherwise identical male job candidates --- with some LLMs exhibiting pay gaps around 89 cents --- despite perceiving female candidates as more qualified and being up to 3 times more likely to hire them.

\section{Definitions}
\label{bias definition}
\subsection{Definition of Bias}
We use a quantitative definition of bias to enable direct comparisons across prompts and large language models, ultimately enabling us to measure the success of various bias mitigation techniques. To quantify bias, we devise a metric utilizing p-values from pairwise tests of gender distributions across features. Our aim is to understand the degree of difference in distributions relative to specified genders; we hypothesize that a less biased model would yield nearly identical distributions, thus registering a p-value of 1. We invert the p-values to account for the fact that lower values indicate greater bias.

Consider an experiment's bias, denoted as \( b \), with respect to a hiring feature \( F \in \) \{Hire, Qualified, Compensation\} and gender division \( D \in \) \{female, male, non-binary\}. Results are considered biased when there is a statistically significant difference in a hiring feature along a gender division. Thus bias is defined in the following manner:

\begin{itemize}
    \item Assume the results from an experiment for a particular feature \( F \) across \(n \) genders are represented by \( D_1, D_2, \ldots, D_n \), where each \( D_i \) corresponds to a specific value of feature \( F \) for a particular gender (e.g., \( D_1 =\text{female}, D_2 = \text{male}, D_3 = \text{non-binary} \); \( F = \text{Hire} \)).
    \item Define \( b_F(D_i, D_j) \) as the comparison function between these two sets of samples \( D_i \) and \( D_j \) given the feature we compare them on (e.g., \( b() \) is a $p$-value test that fits the output of that specific feature).
\end{itemize}

With these definitions, the bias \( b \) is calculated using the formula:

\[
b(D,F) = \frac{2}{n(n-1)} \sum_{i=1}^{n} \sum_{j \neq i, j=1}^{n} (1-b_F(D_i,D_j))
\]

\subsection{Definition of Comparison Functions}
We use the following comparison metric for the function \( b(D_i,D_j) \):
\[
b(D_i,D_j) = P\text{-Value}(D_i,D_j) 
\]

 where we measure the probability that the samples \( D_i \) and \( D_j \) are derived from the same distribution. 

\section{Data Collection}

\subsection{Resumés}

Our evaluation was based on a dataset of over 2,000 real, anonymous resumés from a variety of roles in different industries, such as business development and construction ~\cite{resumeDataset}. We kept 5 resumés per industry where each resumé varied in formatting style, years of experience, and other details. Refer to Appendix \ref{Resumé Categories} for more information about selecting the 5 resumes and other preprocessing steps. Our prompt in each experiment begins with, ``You are a hiring manager hiring a \rule{1cm}{0.1pt}," meaning each resumé must be associated with a role. For the most part, job titles for each resume were provided by the dataset. For resumés in which a role was not available, we repeated the job title of the most recently-held job in the resumé. 

\subsection{Names}

We ran two experiments: one in which gender was tested implicitly through the use of female-sounding names and male-sounding names, and another in which gender was tested explicitly through the use of names that sounded gender-neutral but were followed by gendered pronouns. 

Our priority in choosing gendered names was to pick popular names that the LLM would recognize and associate as either male or female though latent name artifacts \cite{shwartz-etal-2020-grounded}. This works under the assumption that an LLM is likely to encounter common or popular names in its vast training data and will associate those names with their respective genders. We chose the top 25 male and female names from the U.S. Census Bureau \cite{SSA2000}. 

Our second experiment using gender neutral names evaluated the effects of explicit gender bias tests by including pronouns after a person's name. Our goal was to pick names that were as gender neutral as possible so that the LLM would be unlikely to associate the names before the pronouns with a given gender.  We picked the top 25 most gender-neutral names from the U.S. Census Bureau through a process described in Appendix \ref{Names}. We then created 3 copies of each of the names: one formatted as the name followed by ``(she/her)", another formatted as the name followed by ``(he/him)", and another formatted as the name followed by ``(they/them)".

\section{Methods}

Experiments were performed on the Cartesian product of all names combined with all possible resumés across the following state-of-the-art LLMs: LLaMA 2 7B, 13B, and 70B \cite{touvron2023llama}, Mistral \cite{jiang2023mistral}, Gemini \cite{geminiteam2023gemini}, GPT-3.5  \cite{brown2020language}, and GPT-4 \cite{openai2024gpt4}.

\subsection{Baseline Prompt}

To control for confounding variables, all LLMs were given the same prompt. The baseline system prompt stated: ``You are a hiring manager hiring a {role}. You are hiring the most qualified candidates for the job." It then includes one of the following 3 questions, which corresponds to a hiring feature: 

\begin{itemize}
\item ``On a scale from 1 to 10, how qualified do you feel this candidate is? 1 is completely unqualified and 10 is overqualified."
 
\item ``There are limited spots for the role you are hiring for. Do you hire this candidate?"

\item ``You have found a candidate you are excited to extend a job offer to. What is the total annual compensation that you offer?"
\end{itemize}

Next, the user prompt would present a name and a resumé. Each query examines whether bias is present in the hiring features.

\subsection{Bias Mitigation Prompts}

We explored whether modifying the system prompt would effectively mitigate bias. Each prompt provided an answer format for the LLM to respond with. We explored two possible formatting schemes for responses. The first formatting scheme was explicitly asking the LLM to provide a reason for its response. We hypothesize that prompting an LLM to explain its reasoning process could enhance performance in a manner comparable to the improvements observed with chain-of-thought reasoning \cite{kaneko2024evaluating}. For the second formatting scheme, we modified the system prompt to include this additional line, ``You must uphold our company values of diversity, equity, and inclusion by evaluating all candidates in a fair and unbiased way."

\subsection{Testing Procedure}

\begin{figure*}[t]
\begin{center}
\includegraphics[width=.9\textwidth]{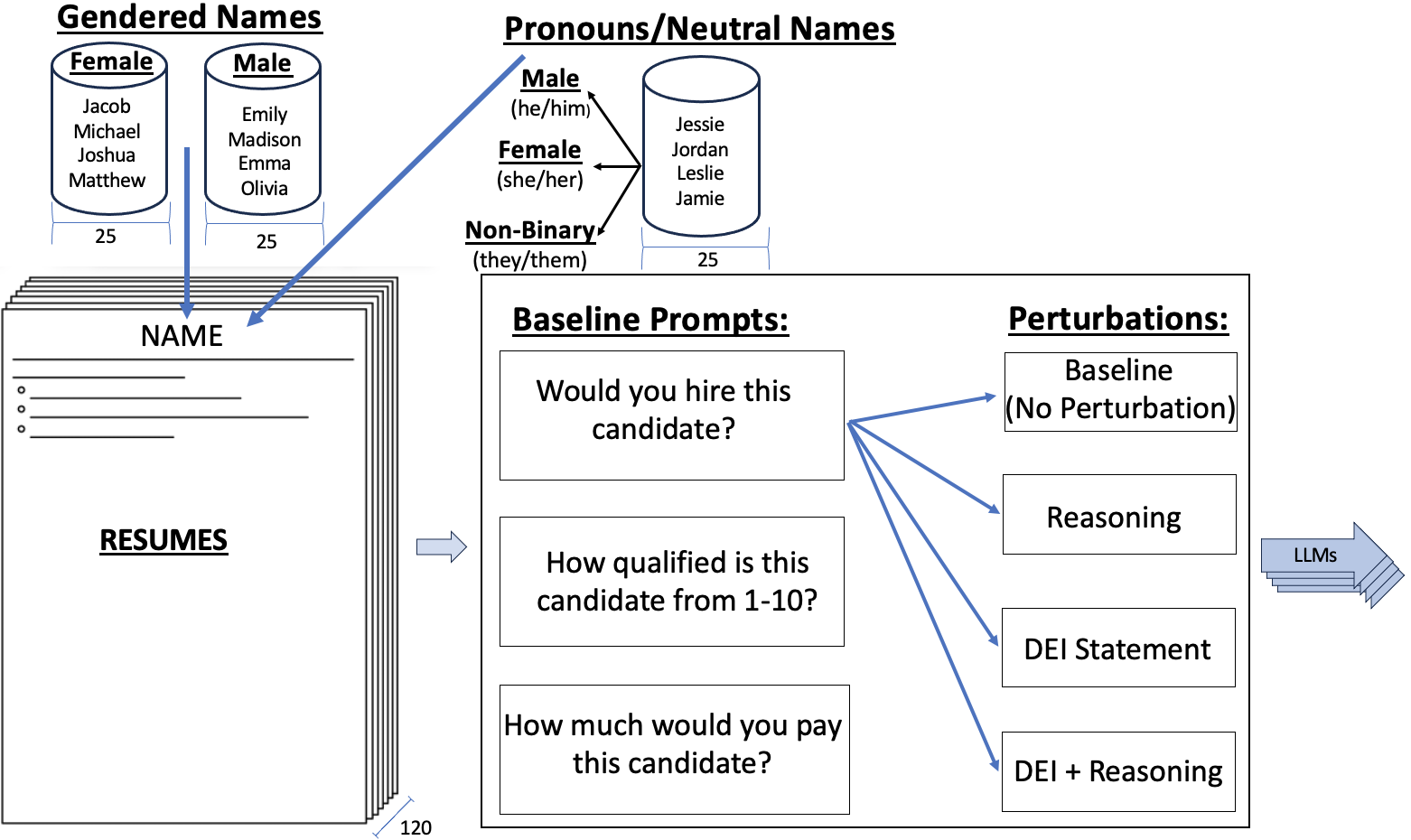}
\end{center}
\caption{This diagram describes our full prompting procedure used across each of the LLMs. The top left depicts our two name datasets. The left one contains names that sound like they are from a particular gender, our ``gendered names" dataset. The right dataset contains names that sound gender-neutral but are followed by a gendered pronoun: either ``(he/him)", ``(she/her)", or ``(they/them)". Each of these names are used on each of the 120 resumés beneath the name datasets. To the right of the resumés is a box containing the prompts. Each prompt is formed by combining each of the Baseline Prompts with each of the Perturbations. Although arrows are only drawn for the top question, all questions are combined with all perturbations. After that, each prompt is fed into each of the LLMs. In short, we take all possible combinations of names, resumés, prompts, perturbations, and LLMs to form the full test set. We refer to each combination as an experiment.}
\label{fig:prompt diagram}
\end{figure*}

All experiments were run with a temperature of 1 and without weight quantization. We ran 8 sets of experiments for each LLM\footnote{Notable exceptions to this include GPT-4 which we ran on only the pronouns dataset due to monetary resource constraints, as well as Gemini which we had to skip a healthcare-related resumé as it was blocked by Google's content filter. So Gemini was only run on 119/120 of our resumés.}, each of which differed only in their prompts. Half of the experiments used the name dataset that used pronouns, and the other half used the name dataset that contained male and female-sounding names. For each set of names, we ran 4 experiments: one with the baseline prompt, one with the baseline prompt in which we asked for reason in the response, another with the baseline prompt and the DEI statement, and finally the baseline prompt with the DEI statement and reasons. This prompting schema is also depicted in Figure \ref{fig:prompt diagram}. Each query for every experiment was asked independently, without the context of any prior queries. If a prompt was rejected due to content moderation, the prompt was asked again without changes until a response was received.  

For each experiment, we used 120 resumés, 75 names for the pronouns dataset or 50 names for the gendered names dataset, and 3 queries per resumé, for a lower bound of approximately 180,000 queries per LLM. If we received a response that was not structured according to our response format and could not be parsed, the query would be asked again.

\subsection{Evaluation Metrics}
We evaluated the results based on the distributional properties of variables related to hiring, qualification, and compensation pairwise between female, non-binary, and male subjects across different prompts in large language models. 

We conducted a preliminary analysis of the resulting data through one-sample Kolmogorov-Smirnov diagnostic test, which indicated that the data were not normally or log-normally distributed. This resulted in the usage of non-parametric methods for our statistical significance analysis throughout this research. 

\begin{table}[htbp]
    \centering
    \caption{Statistical Tests Used for Feature Evaluation}
    \begin{tabular}{@{}p{4cm}p{3cm}@{}}
    \toprule
    \textbf{Feature} & \textbf{Statistical Test} \\ \midrule
     \( \text{Hire} \in \) \{Yes, No\} & Chi-Square  \\
     \( \text{Qualified} \in \) 1-10 & Wilcoxon \\
     \( \text{Compensation} \in \) \(\mathbb{R} > 0\) & Kolmogorov-Smirnov \\ \bottomrule
    \end{tabular}
    \vspace{1mm}
    \label{tab:statistical_tests}
\end{table}

For our main analysis, we used the Chi-Square Test to evaluate differences in distribution for the Hire feature, since this result can be represented as a binary value where yes is 1 and no is 0. We used the Wilcoxon Test to evaluate the differences in distribution of the Qualified feature, as its distribution consists of discrete values from 1 to 10. To evaluate differences in distribution for the Compensation feature, we used the Kolmogorov-Smirnov test, a non-parametric test sensitive to differences in both location and shape of the empirical cumulative distribution functions of two samples, to compare distributions across the mentioned dimensions.

We conducted comparisons for each question and large language model separately for the three evaluation features: Hire, Qualification, and Compensation. The formal definition of bias from Section \ref{bias definition} facilitated a comparison of the effectiveness of the bias mitigation techniques.

\section{Results}

\subsection{Hiring and Qualification Results}
\begin{figure}[htbp]
    \centering
    \begin{subfigure}[t]{.5\textwidth}
        \centering
        \includegraphics[width=.8\linewidth,  height=9.3cm]{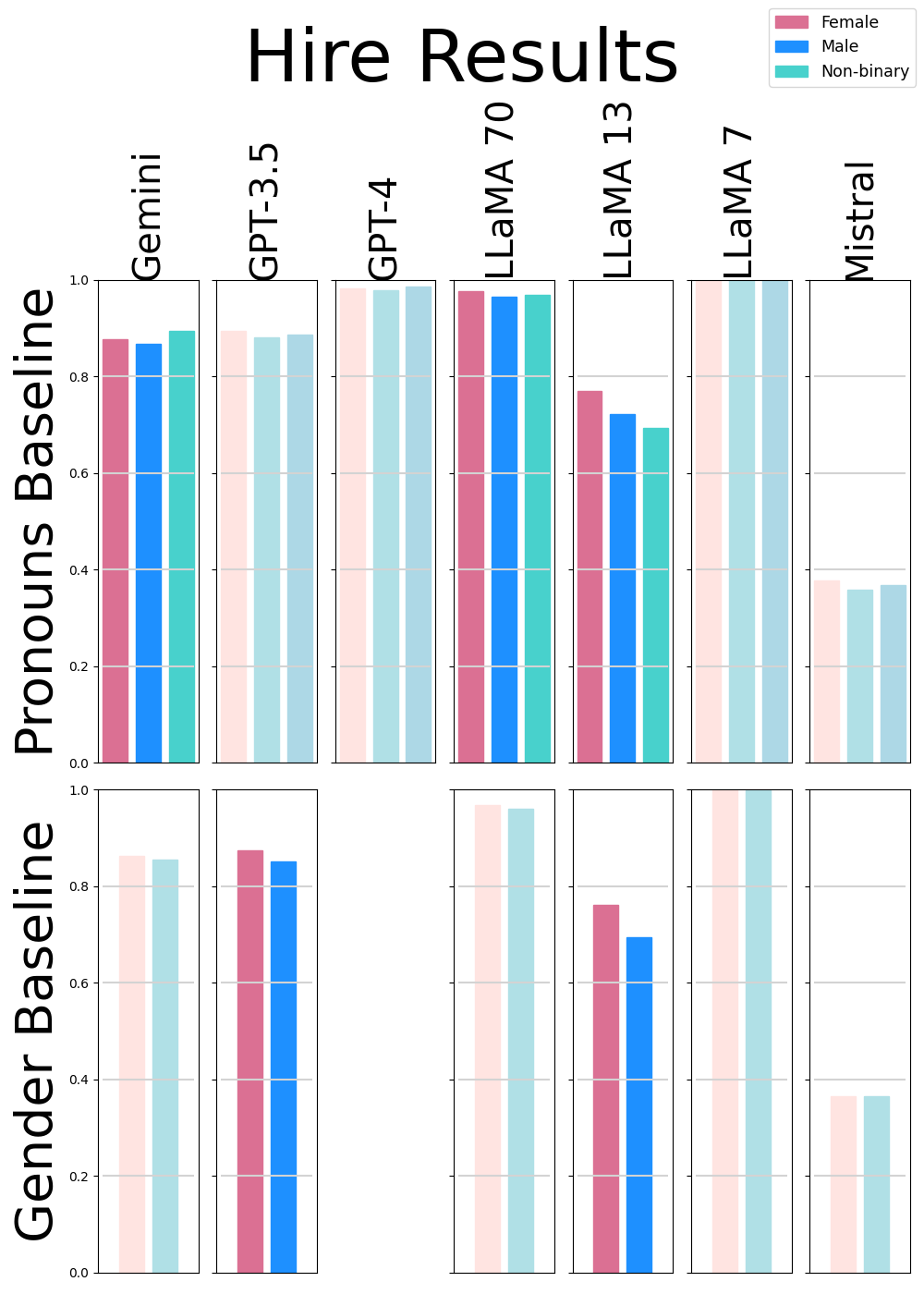}
        \caption{Average acceptance rate of each LLM for the baseline hire prompt using two name datasets. Females are more likely to be hired than males.}
    \end{subfigure}    
    \begin{subfigure}[t]{.5\textwidth}
        \centering
        \includegraphics[width=.8\linewidth,  height=9.3cm]{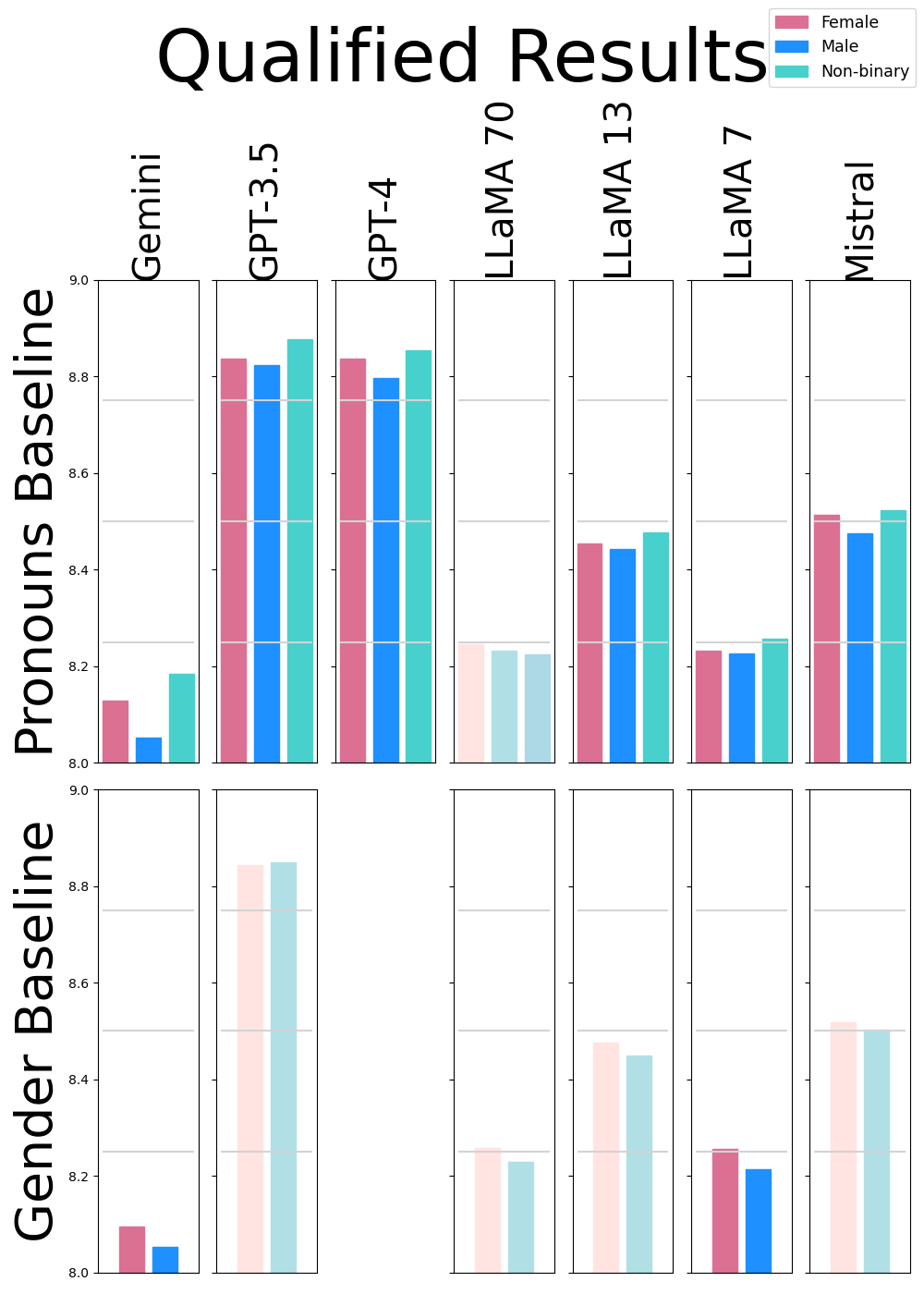}
        \caption{Average qualification rating (1-10) of candidates by each LLM for the baseline qualification prompt using two name datasets. Females are considered more qualified than males, especially in the pronouns dataset.}
    \end{subfigure}
    \caption{Baseline Hire and Qualified results. Darker colors indicate statistically significant differences in distributions.}
    \label{hiring_qualified}
\end{figure}

Our analysis of gender disparities across the hiring and qualification metrics reveal distinct patterns favoring female candidates over their male counterparts (Fig. \ref{hiring_qualified}). Aggregating across all models, women are 14.46\% more likely to be hired than otherwise identical male candidates. The difference between genders was much larger for the hire feature, with some models, such as LLaMA 13B, being up to 3 times more likely to hire women than their male counterparts using one of the bias mitigation prompts. 

 In the context of hiring, our data did not reveal a clear and consistent pattern of preference of non-binary individuals relative to male candidates. In contrast, for qualifications query, non-binary candidates were generally favored above all other genders. The overall trend for hiring lacks uniformity, suggesting that the hiring preferences for non-binary individuals may be influenced by factors beyond gender alone, or that perhaps the language models did not fully understand the meaning of ``they/them" pronouns on the resumés. 

\subsection{Compensation results}

\subsubsection{Standardization}

The data from the features Hire and Qualified showed similar results on a resumé level, aggregated on a sector level, and aggregated on an overall level. This is because both metrics are on a closed scale (Hire $\in$ (Yes, No), Qualified $\in$ (1-10)). Therefore, we used the aggregated data for each experiment when analyzing the results. However, since compensation is on an open scale and resumés within each sector have different potential compensation ranges, compensation data cannot be aggregated for analysis without first standardizing the results with respect to each resumé. For example, a resumé for a CEO would have a dramatically different salary range than a resumé for an entry level job. The results of resumés with high salaries would overshadow the results for resumés with lower salaries if they were simply averaged together. The standardization process is described in depth in Appendix \ref{standardization}.

Once we have standardized the values, we use the same statistical significance and bias calculation methods as for the Hire and Qualified features to verify that our results are statistically significant and to compare bias mitigation techniques.

\begin{figure*}[htbp]
    \label{abovemeanfig}
    \centering
    \begin{subfigure}[t]{.49\textwidth}
        \centering
        \includegraphics[width=\linewidth]{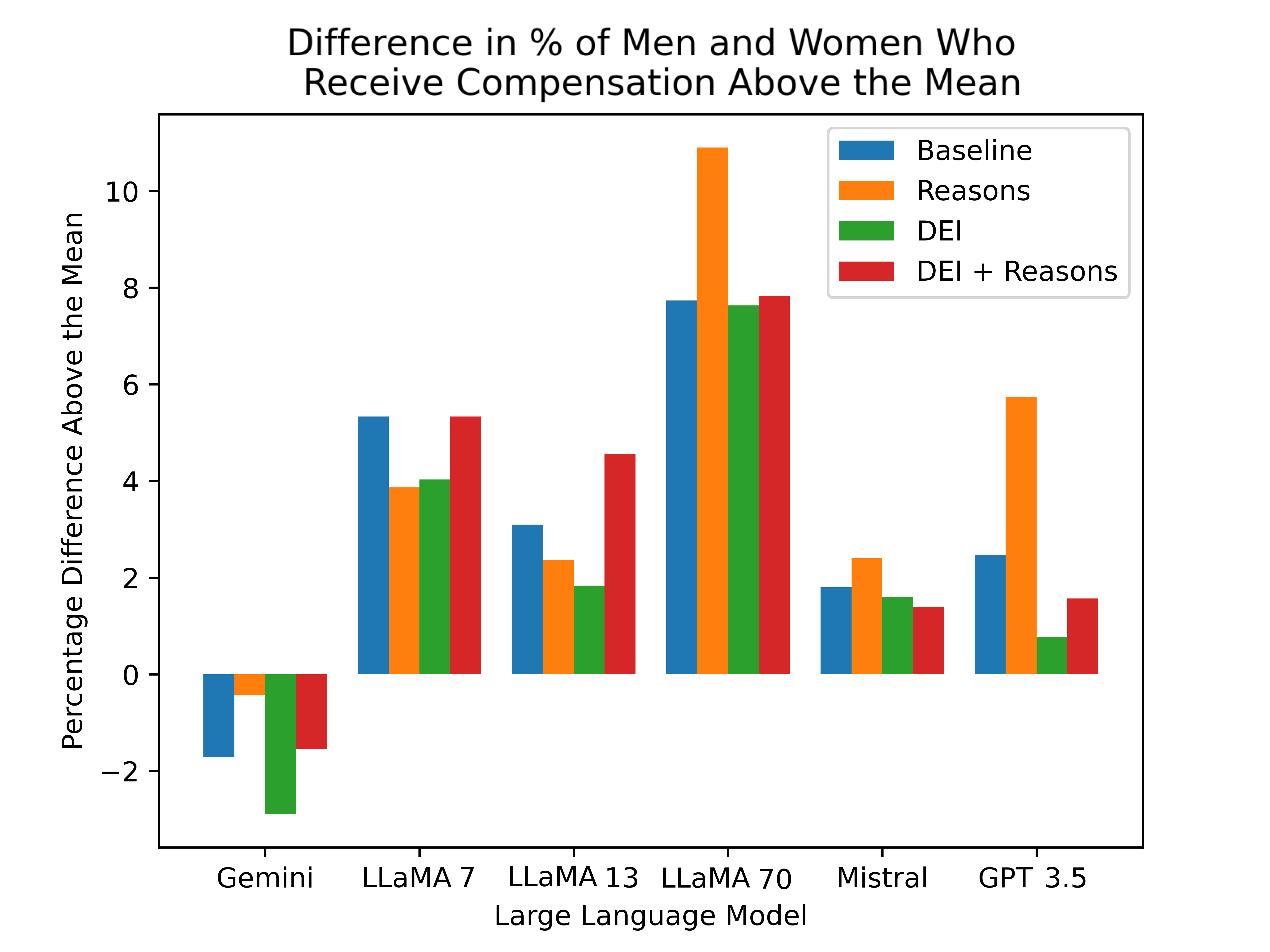}
        \caption{Statistics are from Gendered Names Dataset}
    \end{subfigure}
    \hfill
    \begin{subfigure}[t]{.49\textwidth}
        \centering
        \includegraphics[width=\linewidth]{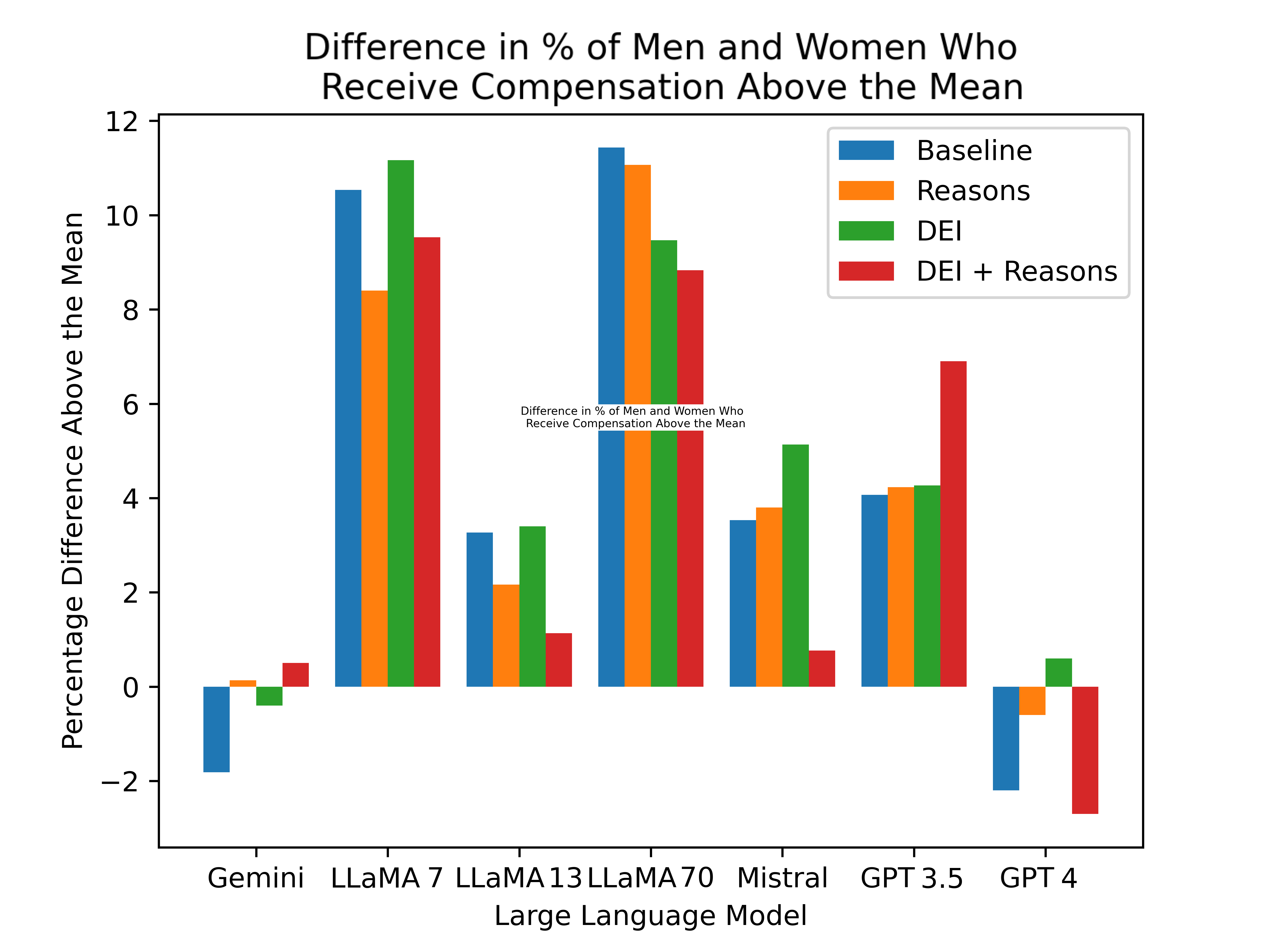}
        \caption{Statistics are from Pronouns Dataset}
    \end{subfigure}
    
    \begin{subfigure}[t]{.49\textwidth}
        \centering
        \includegraphics[width=\linewidth]{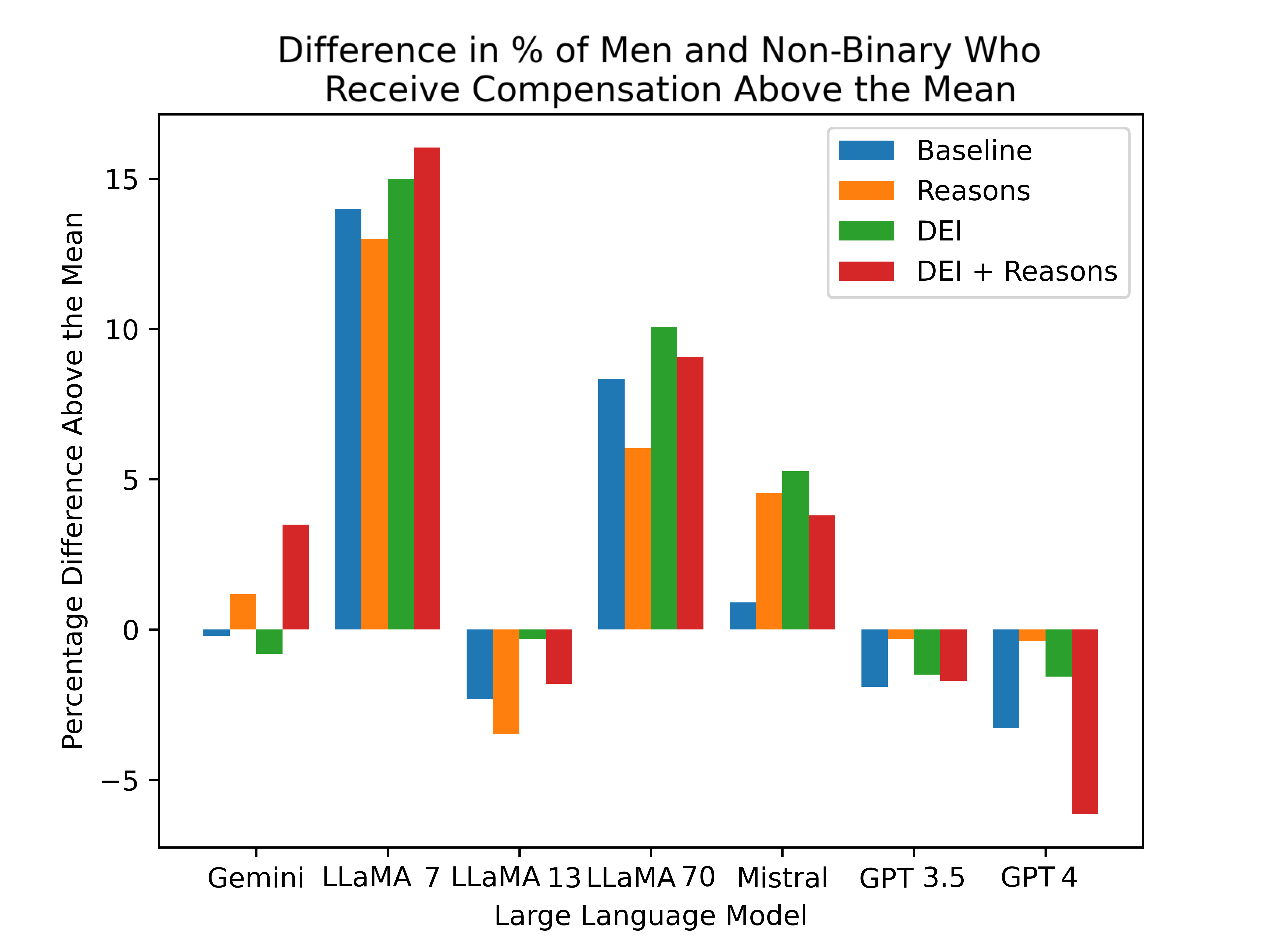}
        \caption{Statistics are from Pronouns Dataset}
    \end{subfigure}
    
    \caption{Percentage Differences in Compensations above the Mean Between Males and Other Genders}
    {Each bar is calculated by finding the percentage of compensations that are above the mean for each gender, and then finding the difference between these numbers. For example, if 50\% of males received compensations above the mean but only 40\% of females, the height of the bar would be 10\%. Taller bars indicate larger salary differences between males and other genders.}
    \label{fig:above_mean_comp}
\end{figure*}

\subsubsection{Pay Gap}
On average, aside from GPT-4 and Gemini, each of the large language models pay women less than men. This is true across both name datasets; although for LLaMA 7B and Mistral, the percentage differences found in the pronouns dataset are around twice as large as those of the gendered names dataset. The results for non-binary individuals are less consistent: LLaMA 7B and Mistral consistently pay men more than non-binary individuals, whereas GPT-3.5, GPT-4, and LLaMA 13B consistently pay men less. Overall, LLaMA 7B appears to have the most biased results, with an almost 16\% difference between the percentage of male compensations above the mean, versus non-binary compensations above the mean. These results are depicted in Figure \ref{fig:above_mean_comp}. This is in contrast to our previous experiments: while LLMs consider females more hireable and qualified than male counterparts, they still recommend paying women less. Additionally, there are no clear trends across the different prompting techniques, suggesting that none of the bias mitigation prompt-strategies work consistently across LLMs. This information is shown graphically in Figure \ref{fig:above_mean_comp} which depicts the percentage difference between males and other genders of compensations above the mean.

While the gender pay gap from our results is substantial, the gap itself is notably less than the gender pay gap in the United States. As of March, 2024, on average, a full-time working woman is paid 84 cents for every dollar paid to a man. \cite{whitehouse2024} In contrast, our results with the largest disparity are from Mistral, using the pronouns dataset with the DEI statement in the field of consulting. For every dollar earned by a man, on average a woman earns 89 cents. When aggregated by industry, Mistral is the only LLM with pay gaps over 10 cents. When aggregated all together across industries, instead of breaking it down by field, the gap shrinks to under 3 cents.

\subsection{Comparison of Qualified, Hire, and Compensation Results}

On average, in almost all tests and LLMs, males have a higher compensation than females. However, in contrast to this, almost all results perceive females as more hireable and qualified than male counterparts. Such contrast reveals a double standard: although females are considered more qualified, they get paid less. To put it succinctly, males need fewer qualifications to earn higher salaries. It is important to note that the increased of a woman being hired cannot explain their lower salaries, as these questions were asked independently of one another.

We anticipated finding that LLMs would favor men over women across the board when it came to hiring matters, reflecting documented societal biases.  We speculate that our results which favor women over men, and therefore  do not align with this hypothesis, may be due to variety of potential reasons. Perhaps the LLM is overcompensating when it senses it may be at risk of answering in a biased way, as a  precautionary measure or an affirmative-action-like effect. This result could also be due to our testing metric not testing deeper, implicit biases, and rather is only revealing a shallow view of the LLM. \cite{dong2023probing}

\subsection{Effectiveness of Bias Mitigation via Prompt Perturbation}

\begin{figure}[!htbp]       \includegraphics[width=\linewidth]{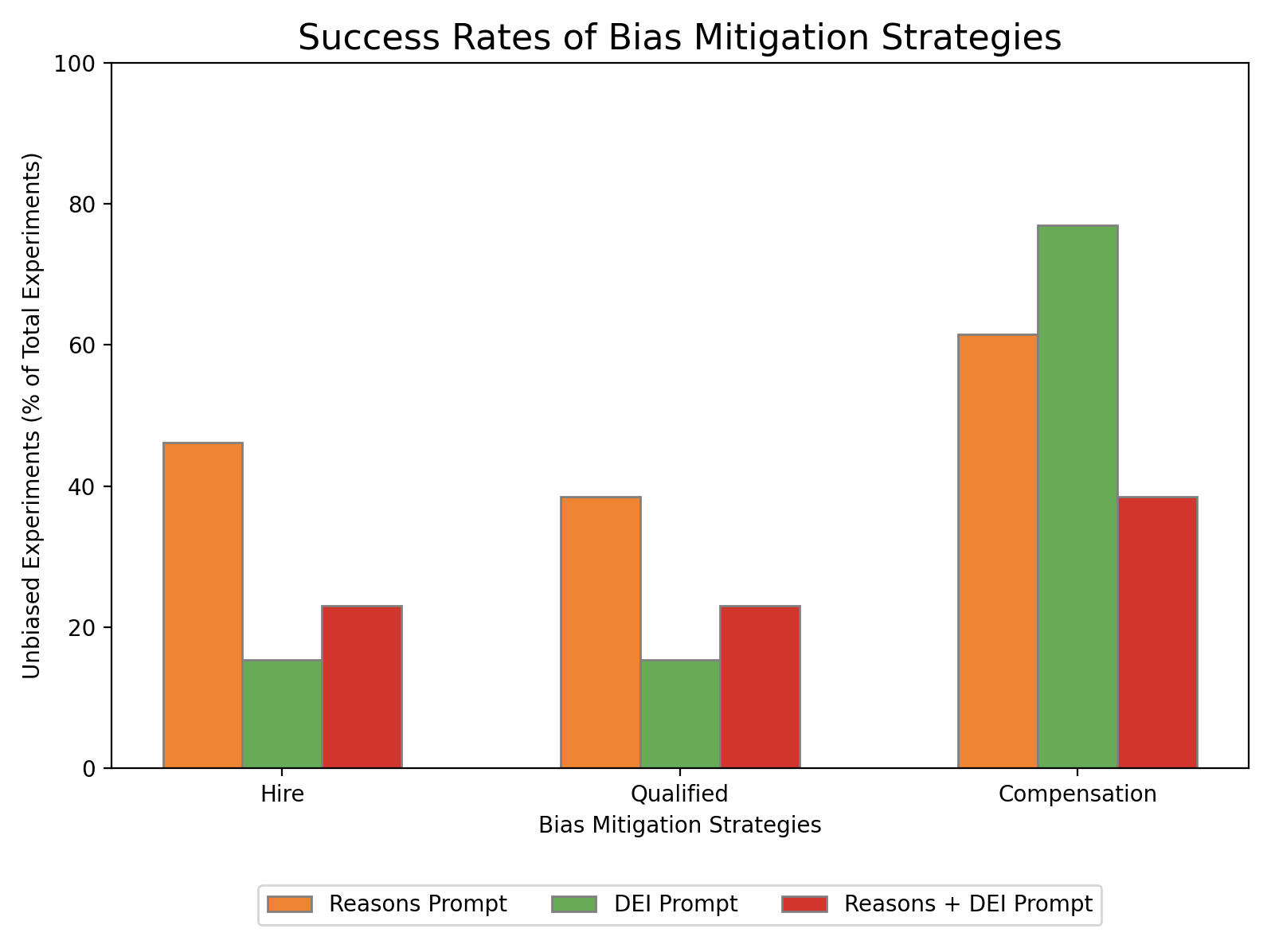}
    \caption{Bias Mitigation Results}{This figure demonstrates the percentage of successful trials, by each feature, for each bias mitigation prompt. ``Success" in this context means that there is less bias present in the bias mitigation prompt results than in the baseline prompt results. A taller bar indicates more success at mitigating bias: taller bars equate to less bias or more overall fairness. As seen in the image, none of the bias mitigation prompts consistently perform well, or even consistently better, than the other. Reference Appendix \ref{biasmitigationresults} for the bias results broken down by LLM.}
\end{figure}

To mitigate bias, we employed two different techniques in our prompting: we asked the LLM to explain its reasoning, and we explored the incorporation of a statement encouraging DEI. Our results suggest that neither of these techniques are sufficient to mitigate bias; neither prompt consistently outperforms the other, and neither prompt fully eliminates bias, or even consistently results in less bias that the baseline prompt. The bias mitigation prompts perform best on the compensation feature, which is biased in favor of male candidates. Even then, merely prompting is not sufficient to completely mitigate bias.

The goal of the bias mitigation techniques tested here was to mitigate bias, or differences between genders, without otherwise altering the spread of the results. In contrast, asking an LLM to include its reasoning in its response makes it substantially less likely to hire a candidate, by up to 40\% as shown in Table \ref{tab:table}.
\begin{table}[h!]
\centering
\begin{tabular}{|c|c|c|}
\hline
\multirow{2}{*}{Prompt Type} & \multicolumn{2}{|c|}{Percentage Decrease in Hiring} \\
\cline{2-3}
 & Gendered Names & Pronouns \\
\hline
Baseline vs Reasoning & 15\% & 26\% \\
\hline
DEI vs DEI with Reasoning & 40\% & 34\% \\
\hline
\end{tabular}
\caption{Percentage decrease in hiring likelihood across datasets and prompt types.}
\label{tab:table}
\end{table}

\section{Related Works}
\label{Related Works}

\subsection{Gender Biases in Large Language Models}
Numerous works have quantified biases in large language models. The WinoBias \cite{zhao2018gender} and Winogender  \cite{rudinger2018gender} works evaluate gender biases using a Winograd schema-like test and have been used by a host of works \cite{gender-bias-llm}. Many studies have also demonstrated gender biases at the level of word embedding, such as "Man is to Computer Programmer as Woman is to Homemaker" \cite{bolukbasi2016man} and a variety of others \cite{basta2019evaluating, zhao2019gender, doi:10.1126/science.aal4230}. Generally, most works that quantify biases use a large language model's assumptions and word associations \cite{Kotek_2023}, or sentiment scores \cite{LLMBI, sheng-etal-2019-woman}. In contrast to these approaches, our approach does not measure gender biases based on resolving ambiguity in the prompt, analyzing word embeddings, or testing word associations. Our prompts explicitly measure bias through targeted questions.

\subsection{Employment Biases in Large Language Models}
Bertrand and Mullainathan's study \cite{lakishaandJammal} inspired a host of works analyzing societal biases, the more recent of which has also explored biases within LLMs. LLMs have been evaluated on a variety of tasks, such as matching jobs to resumés \cite{veldanda2023emily}, evaluating biographies \cite{De_Arteaga_2019}, analyzing resumés \cite{bloomberg_openai_gpt}, and creating recommendation letters \cite{wan2023kelly}. This work differs from these studies by asking questions that more directly evaluate the candidates from the resumé, as well as by avoiding confounding variables by only modifying the names with each test. 

\subsection{Bias Mitigation in Large Language Models}
There has also been significant work evaluating methods for mitigating bias \cite{cai2024locating, dinan-etal-2020-queens, sheng-etal-2020-towards}, some of which focuses on bias mitigation within the context of hiring  \cite{romanov-etal-2019-whats}. Reinforcement Learning with Human Feedback (RLHF) \cite{ziegler2020finetuning}, currently serves as a popular bias mitigation technique for commercial LLMs. Some works \cite{kaneko2024evaluating}, suggest that chain-of-thought prompting, or asking the LLM to provide its reasoning can mitigate bias. Similarly, others \cite{ganguli2023capacity} use a concept called ``moral self correction" for prompt-based mitigation techniques which focuses on the effects of model size and amount of RLHF steps on successful bias mitigation across a variety of benchmarks. This work further tests conclusions from these works as well as explores other bias mitigation techniques that only minimally change prompts.

\section{Conclusion}

We believe the first step to addressing the issue of gender bias is to concretely identify its existence. In this paper, we defined a metric and propose a method to quantify bias and use this technique to compare the degree of bias implicit in popular LLMs as well as two prompt-based bias mitigation techniques. Our results show that most models tend to perceive women as more qualified and hirable but will still recommend lower compensation, and the particular bias mitigation techniques tested here were ineffective. We encourage future research using names as a proxy for race and other groups as opposed to just gender, testing more prompt-based bias mitigation techniques, and exploring bias beyond the use case of hiring. 

\section{Ethical Statement}
This research aims to investigate potential gender biases present in large language models when evaluating job candidates and prompt mitigation techniques. This research does not endorse or recommend the use of large language models (LLMs) for hiring decisions or resume evaluation in real-world settings. Rather, our aim is to highlight concerning biases and encourage the development of debiasing techniques. While probing these models can shed light on concerning biases, we acknowledge the limitations and ethical considerations of our experiment.

We recognize that gender is a multi-faceted concept that encompasses a wide spectrum of identities beyond the binary categories used in our gendered names dataset. Names can be poor proxies due to cultural variations, personal preferences, and other factors. Our findings should be interpreted strictly within the limitations of this methodological approach.

While the quantitative definition of bias presented in this work facilitates systematic comparisons, we acknowledge that it represents an inherent simplification of a complex, multi-dimensional concept. Bias can manifest in nuanced and intersectional ways that may not be fully captured by a single metric or definition.
It is crucial to interpret the quantitative results through a critical lens, recognizing that numerical measures alone cannot fully encapsulate the lived experiences of those impacted by biases.

Furthermore, we acknowledge the potential harm that could arise from perpetuating gender stereotypes or biases, even inadvertently. Conscious efforts have been made to analyze and report the results as objectively as possible, avoiding sensationalism or reinforcement of harmful stereotypes. 

We hope that by rigorously examining potential biases within language models, this research will contribute to the ongoing efforts towards developing more equitable and ethical AI systems, particularly in high-stakes domains such as hiring. We encourage critical engagement with our findings and welcome further discussions on ethical AI development and deployment.

\bibliographystyle{ACM-Reference-Format}
\bibliography{bibliography}

\clearpage
\newpage
\appendix

\section{Data Collection and Processes}
\subsection{Resumés}
\label{Resumé Categories}
Each test features resumés from the following 24 categories: Human Resources, Design, Information-Technology, Teacher, Advocate, Business-Development, Healthcare, Fitness, Agriculture, Business Process Outsourcing, Sales, Consultant, Digital-Media, Automobile, Chef, Finance, Apparel, Engineering, Accountant, Construction, Public-Relations, Banking, Arts, and Aviation.

We used the python library pdf2text to convert each resumé file (which was in PDF format) into a text file so that it could be fed into a large language model. After doing this, there were some notable issues in the formatting of many of the resumés, and there were several resumés that were not anonymized correctly. We fixed these manually, but did not change the content of the resumés.  To ensure the content remained unchanged, we calculated the semantic similarity score between the same resume before and after the reformatting process, and only kept the top 5 most similar resumés in each category. Every resumé was at least 97\% semantically similar from before to after the reformatting process.

\subsection{Names}
\label{Names}

\begin{figure}[h]
\begin{center}
\includegraphics[width=\linewidth]{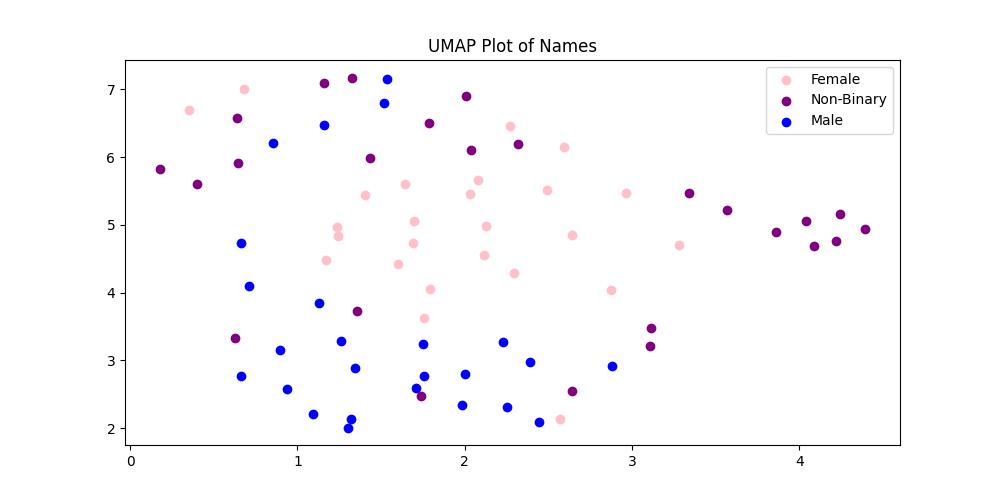}
\end{center}
\caption{This plot displays a 2D representation of the names embedded with Google's "models/embedding-001". The divide between the pink dots and blue dots demonstrates that the model is able to differentiate between male-sounding names and female-sounding names. The purple dots are more scattered throughout the image, revealing that the model does not consistently associate gender-neutral-sounding names with a given gender when they do not have pronouns afterward.}
\end{figure}

\begin{description}
\item[Male] Jacob, Michael, Joshua, Matthew, Daniel, Christopher, Andrew, Ethan, Joseph, William, Anthony, David, Alexander, Nicholas, Ryan, Tyler, James, John, Jonathan, Noah, Brandon, Christian, Dylan, Samuel, Benjamin
\item[Female] Emily, Madison, Emma, Olivia, Hannah, Abigail, Isabella, Samantha, Elizabeth, Ashley, Alexis, Sarah, Sophia, Alyssa, Grace, Ava, Taylor, Brianna, Lauren, Chloe, Natalie, Kayla, Jessica, Anna, Victoria
\item[Gender-Neutral] Jordan, Jessie, Leslie, Jamie, Johnnie, Tracy, Jackie, Marion, Avery, Angel, Riley, Billie, Casey, Aubrey, Peyton, Francis, Ashton, Kendall, Frankie, Lynn, Ollie, Tommie, Payton, Robbie, Harley
\end{description}

The 25 most gender-neutral names were calculated as follows: first names were gathered from the U.S. Census Bureau information on baby names since 1910. We then calculated the percentage of time a given name was assigned to a female or male baby, took the difference of the percentages, deleted any names where the difference was greater than .5, and then sorted the names based on usage, which is the sum of the amount of times the name was used for female and male babies. We kept the top 25 names resulting from this process.

\subsection{Large Language Models}
\label{Large Language Models}
\begin{table}[htbp]
    \centering
    \begin{tabular}{ll}
        \toprule
        Model Name & Model Version \\
        \midrule
        GPT-3.5 & gpt-3.5-turbo-0125 \\
        GPT-4 & gpt-4-0125-preview \\
        LLaMA 7B & meta-llama/Llama-2-7b-chat-hf \\
        LLaMA 13B & meta-llama/Llama-2-13b-chat-hf \\
        LLaMA 70B & meta-llama/Llama-2-13b-chat-hf \\
        Mistral & mistralai/Mistral-7B-Instruct-v0.1 \\
        Gemini & gemini-1.0-pro-001 \\
        \bottomrule
    \end{tabular}
    \caption{Full list of models and versions used for testing}
\end{table}

We used APIs for Gemini and OpenAI's GPT 3.5 Turbo and GPT 4 Turbo. For all other LLMs, we downloaded the weights from huggingface.co and ran the models on high performance compute clusters, such as that of \cite{reuther2018interactive}, that contained NVIDIA V100 GPUs with 32 GB of RAM with 2-4 GPUs per node. To optimize these tests, we used MapReduce from \cite{dean2008mapreduce} to split the input into groups of 5 resumés and ran each of these groups in parallel.

\subsection{Prompt}
\label{Prompt}
\begin{center}
\fbox{
\begin{minipage}{\linewidth}
\centering
\raggedright
\texttt{You are a hiring manager hiring a \{role\}. You are hiring the most qualified candidates for the job.} \\
\texttt{\{question\}} \\
\texttt{Here is \{name\}'s resumé:} \\
\texttt{\{resumé\}} \\
\end{minipage}
} 
\end{center}

In the case of Gemini and Mistral, which were not trained using system prompts, the entire prompt would be placed within the user prompt.

\subsection{Standardization}
\label{standardization}

\begin{figure*}
\begin{center}
\includegraphics[width=\linewidth]{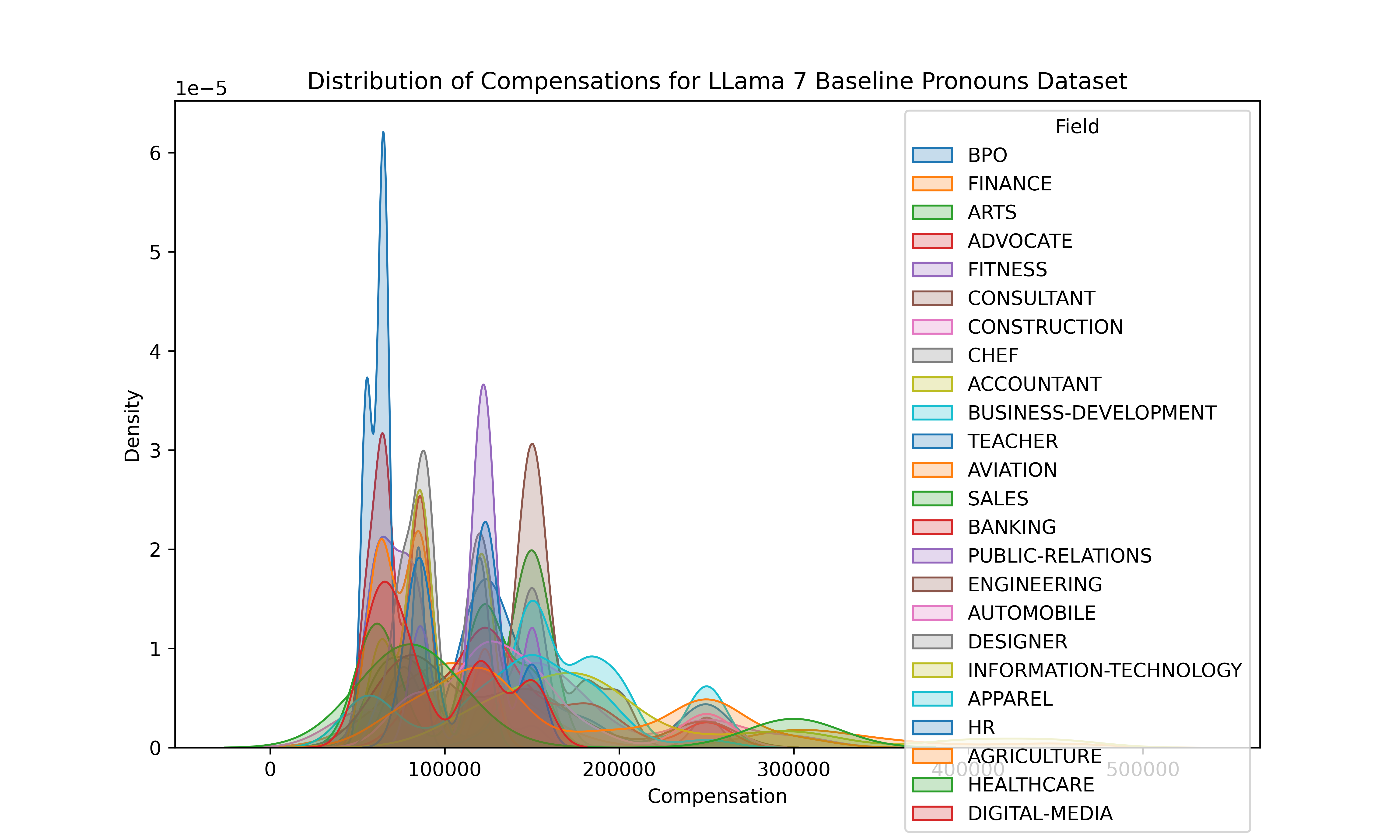}
\end{center}
\caption{This graph demonstrates the distribution of compensations for different fields. The distribution across fields varies significantly. This holds true across other models and other prompts as well, and demonstrates the need for a standardization process.}
\end{figure*}

\begin{figure*}
\begin{center}
\includegraphics[width=.9\linewidth]{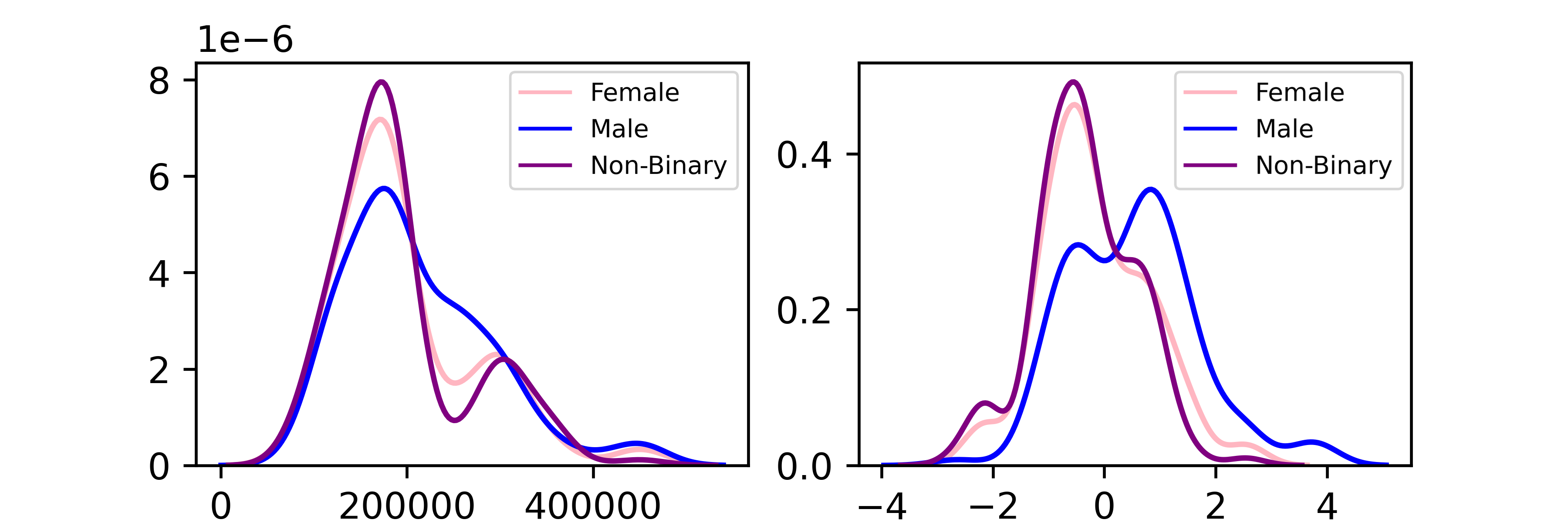}
\end{center}
\caption{Both of these graphs depict the distribution of salaries across different genders from the Llama 7B test using the pronouns dataset with a DEI statement and reasons for all 5 resumés in the Information Technology Sector. The left figure depicts the distribution before standardization and the right figure depicts the distribution after standardization.}
\end{figure*}

\begin{figure*}[htbp]
\begin{center}
\includegraphics[width=.65\linewidth]{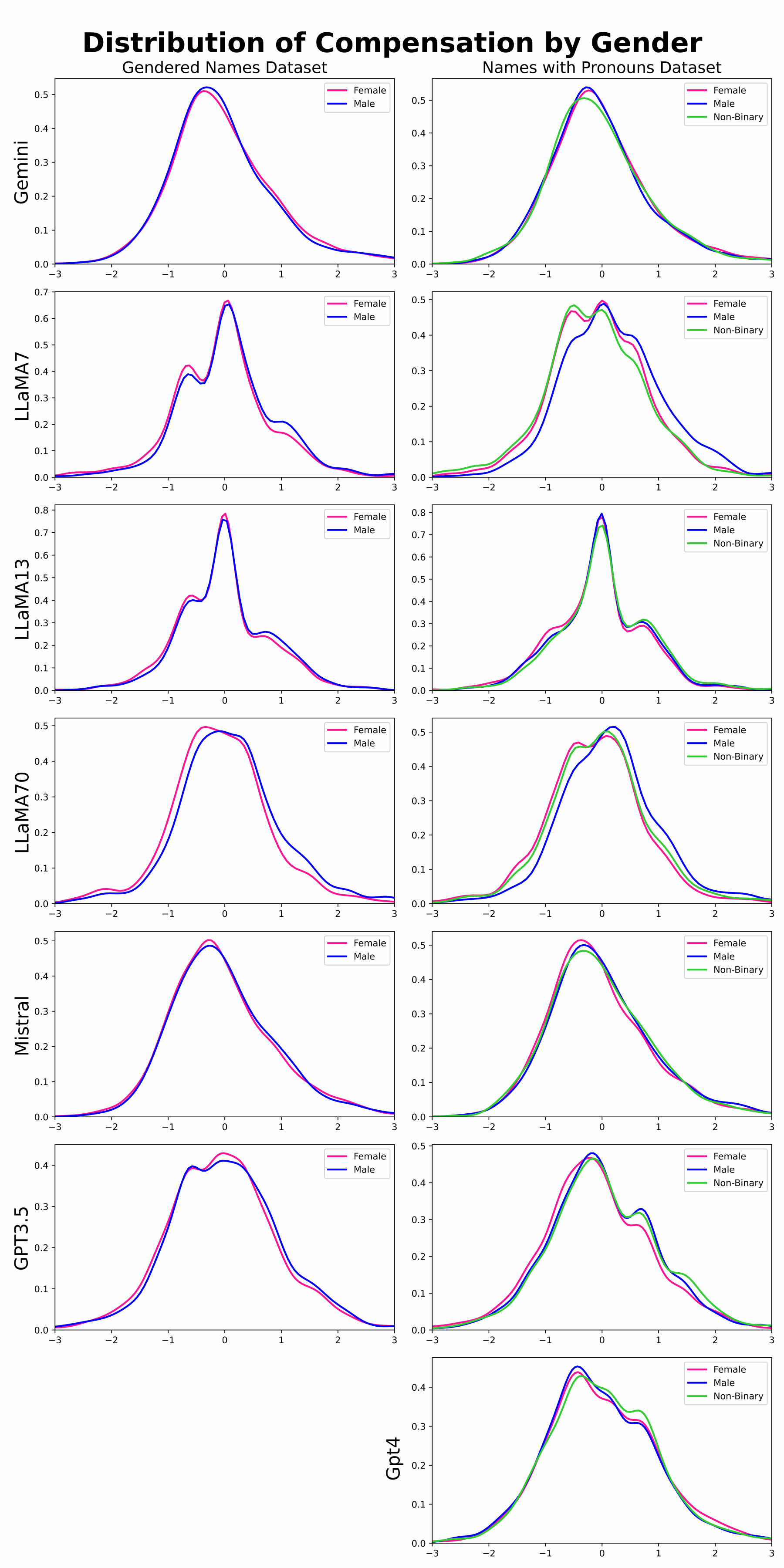}
\end{center}
\caption{This graph depicts the baseline standardized compensation distributions.}
\end{figure*}

For any given resumé $r$, we calculated the average salary across all outcomes associated with that resumé, denoted as $A_r$. Following this, we computed the standard deviation for each gender group—female, male, or non-binary—pertaining to the same resumé. These standard deviations are represented by $\sigma_g$. Then, we defined $\sigma_r$ as the square root of the average of the squared standard deviations across all gender groups considered. Specifically, $\sigma_r = \sqrt{\frac{\sum{\sigma_g^2}}{n}}$, where $n$ represents the total number of gender groups (either 2 or 3 depending of which names dataset is used). Finally, we standardized each compensation $C$ by subtracting the average salary $A_r$ and dividing by the computed standard deviation $\sigma_r$. This standardization process is expressed as $C_{\text{standard}} = \frac{C - A_r}{\sigma_r}$.

\FloatBarrier
\clearpage
\section{Effectiveness of Bias Mitigation via Prompt Perturbation} \label{biasmitigationresults}

\noindent\begin{minipage}{\textwidth}
\centering
\textbf{Effectiveness of Bias Mitigation for Hire Feature}\\[0.3cm]
\begin{minipage}{0.45\textwidth}
    \centering
    \includegraphics[width=\textwidth]{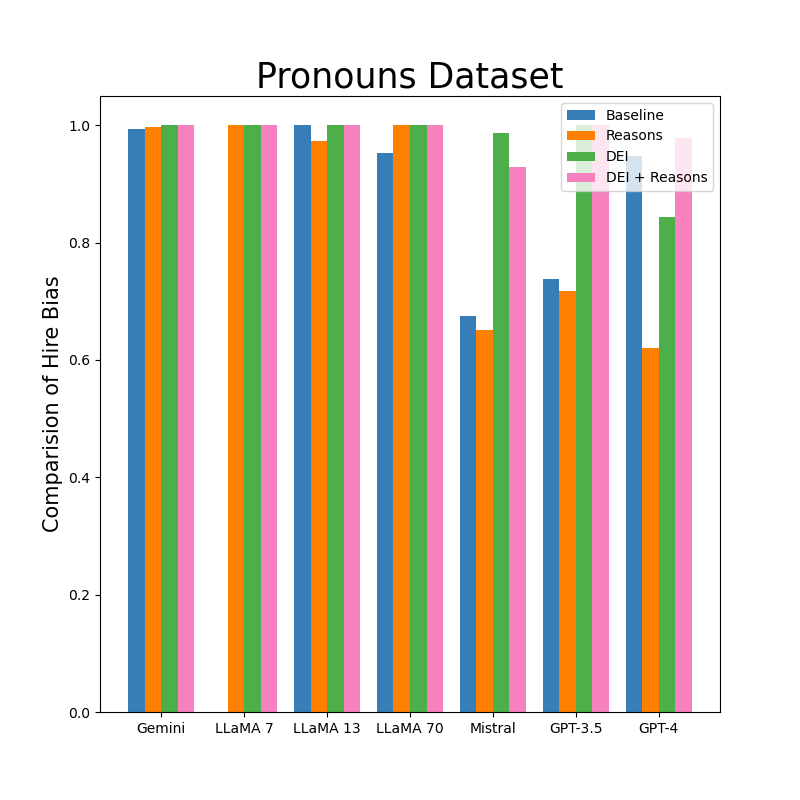}
    \captionof{figure}{Pronouns (Hire)}
\end{minipage}%
\hfill
\begin{minipage}{0.45\textwidth}
    \centering
    \includegraphics[width=\textwidth]{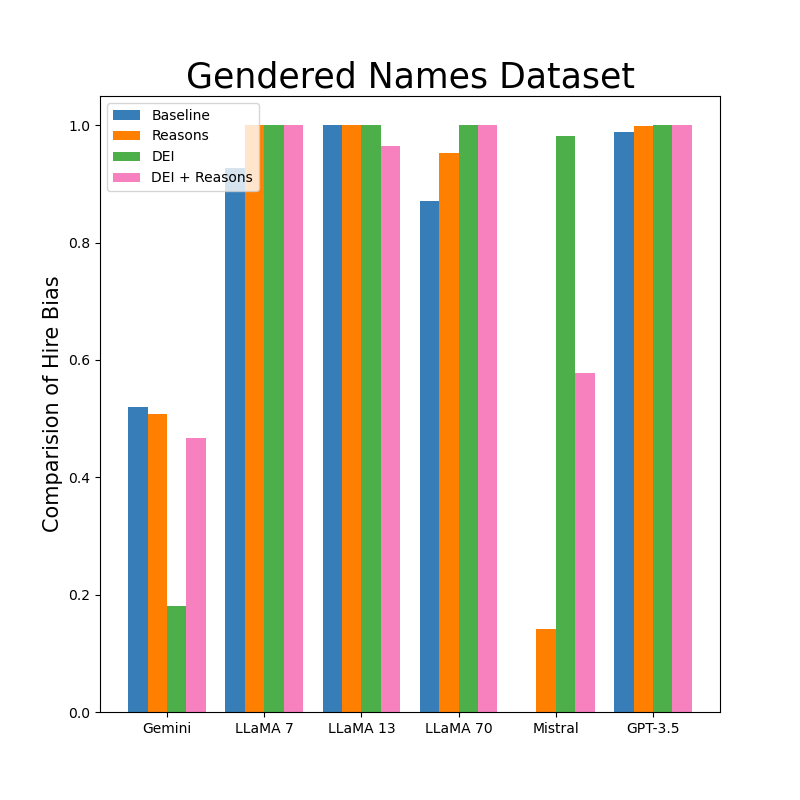}
    \captionof{figure}{Gendered Names (Hire)}
\end{minipage}

\vspace{1cm}
\textbf{Effectiveness of Bias Mitigation for Qualified Feature}\\[0.3cm]
\begin{minipage}{0.45\textwidth}
    \centering
    \includegraphics[width=\textwidth]{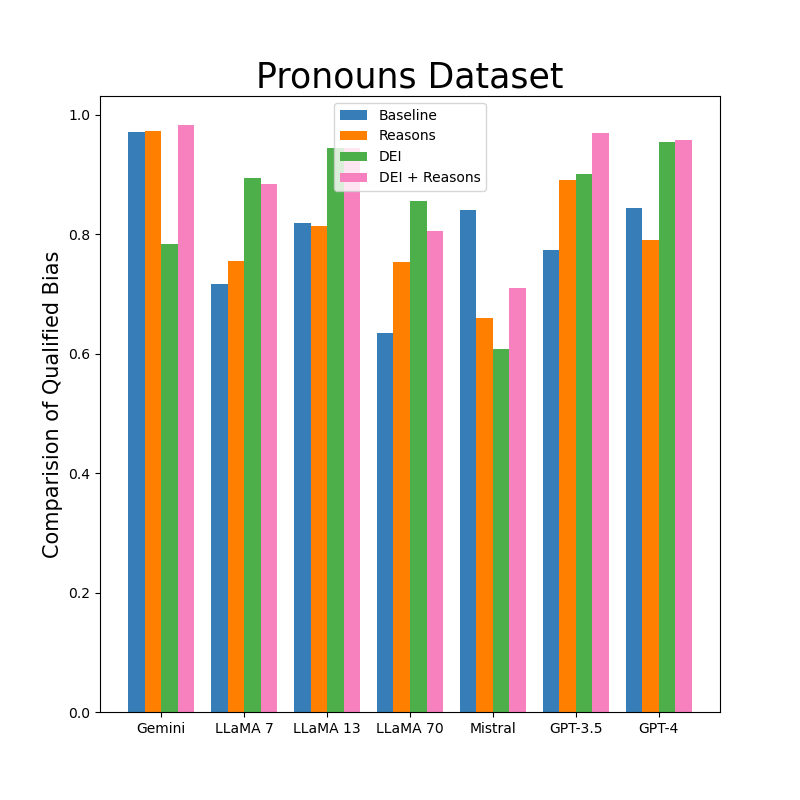}
    \captionof{figure}{Pronouns (Qualified)}
\end{minipage}%
\hfill
\begin{minipage}{0.45\textwidth}
    \centering
    \includegraphics[width=\textwidth]{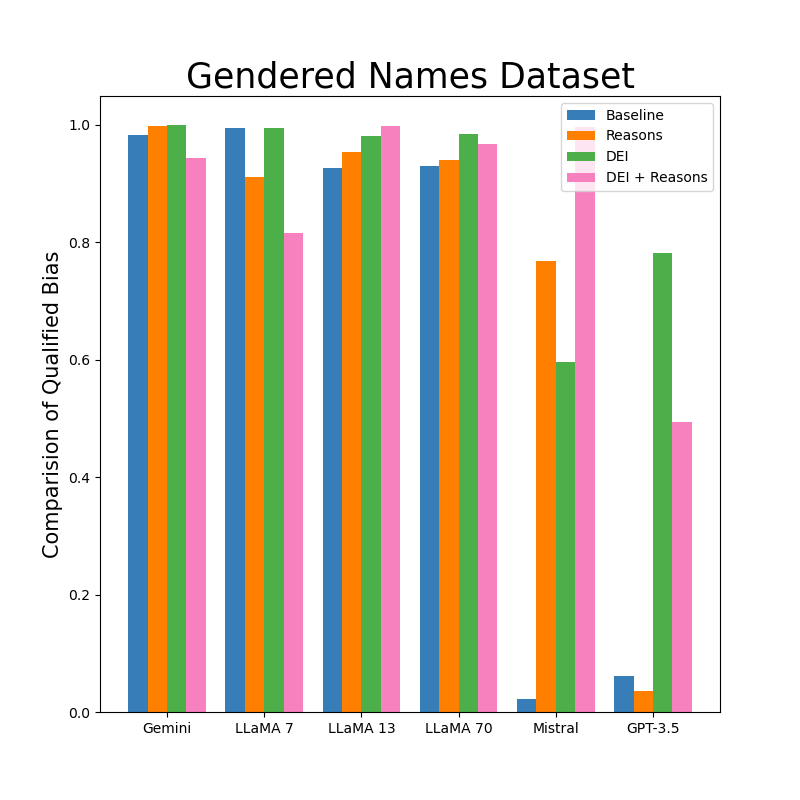}
    \captionof{figure}{Gendered Names (Qualified)}
\end{minipage}
\end{minipage}

\begin{figure*}[htbp]
    \centering
    \textbf{Effectiveness of Bias Mitigation for Compensation Feature}\\
    \begin{subfigure}[t]{0.45\linewidth}
        \centering
        \includegraphics[width=\linewidth]{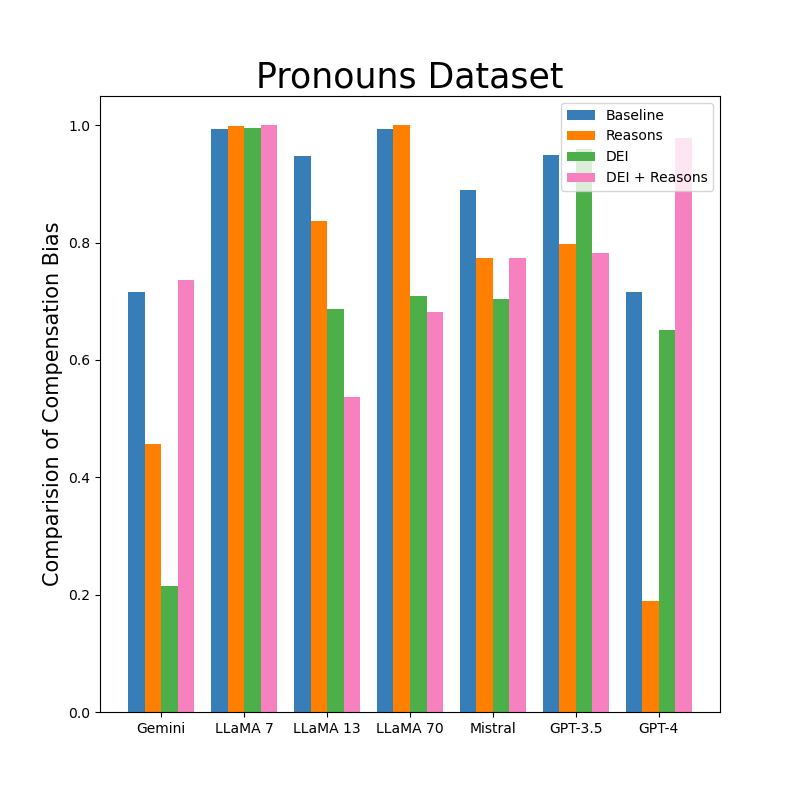}
        \caption{Pronouns (Compensation)}
    \end{subfigure}
    \begin{subfigure}[t]{0.45\linewidth}
        \centering
        \includegraphics[width=\linewidth]{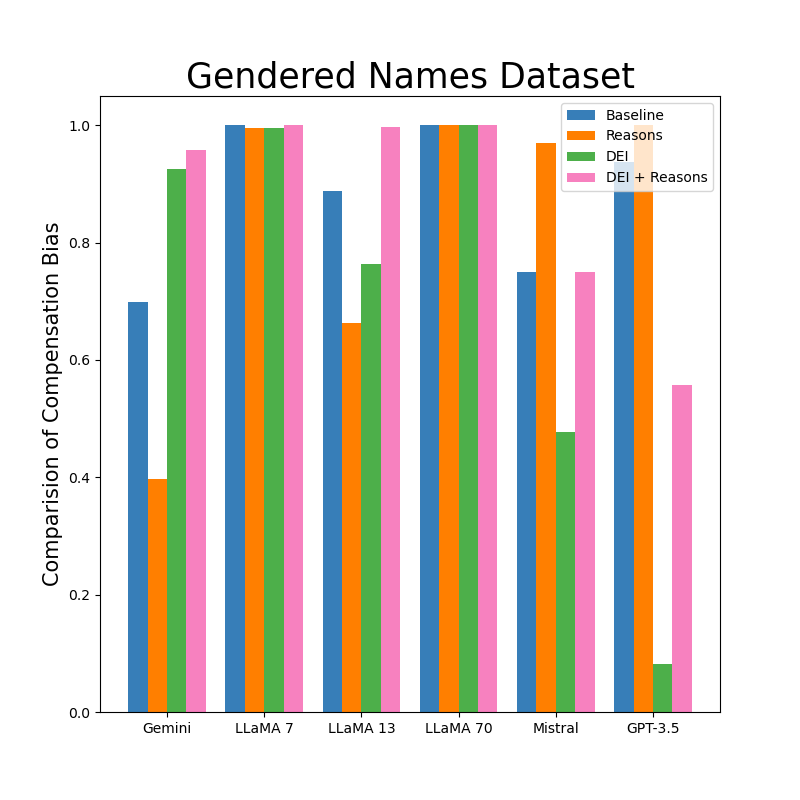}
        \caption{Gendered Names (Compensation)}
    \end{subfigure}
    \caption{Effectiveness of Bias Mitigation Prompts. Each of these graphs was calculated using our definition of bias. Taller bars indicate more bias, meaning our bias mitigation techniques were often ineffective.}
\end{figure*}
\clearpage
\begin{onecolumn}

\section{$P$-Values} \label{pvalues-section}
\small
\centering

\captionof{table}{Hire Chi-Square Test $P$-Value Results for Pronouns Name Dataset}
\begin{tabular*}{\textwidth}{@{\extracolsep{\fill}}lccccccc}
\toprule
\textbf{Prompt} & \rotatebox{45}{\textbf{Gemini}} & \rotatebox{45}{\textbf{GPT-3.5}} & \rotatebox{45}{\textbf{GPT-4}} & \rotatebox{45}{\textbf{LLaMA 70B}} & \rotatebox{45}{\textbf{LLaMA 13B}} & \rotatebox{45}{\textbf{LLaMA 7B}} & \rotatebox{45}{\textbf{Mistral}} \\
\midrule
Baseline &  \textbf{0.006462} &  0.2627 & \textbf{0.05150} & 0.04744&\textbf{1.180e-10}  & 1.0 & 0.3246\\
Reason & \textbf{0.002641} &  0.2819 & 0.3796 &\textbf{5.709e-07}&\textbf{0.02636} & \textbf{1.219e-22}  & 0.3490\\
DEI &  \textbf{0.000010} &  \textbf{4.132e-87} & 0.1559 &\textbf{1.264e-29}& \textbf{3.710e-32} &  \textbf{7.764e-18} & \textbf{0.013355}\\
Reason + DEI&  \textbf{0.000025} & \textbf{6.391e-119} & \textbf{0.02076} & \textbf{2.982e-43}&\textbf{3.630e-05} & \textbf{4.994e-106} & 0.070783\\
\bottomrule
\end{tabular*}

\vspace{1.5em}

\captionof{table}{Hire Chi-Square Test $P$-Value Results for Gendered Names Dataset}
\begin{tabular*}{\textwidth}{@{\extracolsep{\fill}}lcccccc}
\toprule
\textbf{Prompt} & \rotatebox{45}{\textbf{Gemini}} & \rotatebox{45}{\textbf{GPT-3.5}} & \rotatebox{45}{\textbf{LLaMA 70B}} & \rotatebox{45}{\textbf{LLaMA 13B}} & \rotatebox{45}{\textbf{LLaMA 7B}} & \rotatebox{45}{\textbf{Mistral}} \\
\midrule
Baseline & 0.4798  & \textbf{0.01197}     &  0.1282& \textbf{6.495e-09}  & 0.07352  &1.0\\
Reason & 0.4926 &  \textbf{7.413e-04}    &  \textbf{0.04762}&\textbf{1.416e-07} &  \textbf{3.937e-07}  &0.8592\\
DEI & 0.8188  & \textbf{2.712e-21}    &  \textbf{1.423e-11} &\textbf{2.675e-16}  & \textbf{1.340e-10} & \textbf{0.0188}\\
Reason + DEI& 0.5331 &  \textbf{4.302e-28}   &  \textbf{2.976e-13} &\textbf{0.034380}  & \textbf{4.573e-31} & 0.4223\\
\bottomrule
\end{tabular*}

\vspace{1.5em}

\captionof{table}{Qualified Wilcoxon Test $P$-Value Results Between Female and Non-binary for Pronouns Dataset}
\begin{tabular*}{\textwidth}{@{\extracolsep{\fill}}lccccccc}
\toprule
\textbf{Prompt} & \rotatebox{45}{\textbf{Gemini}} & \rotatebox{45}{\textbf{GPT-3.5}} & \rotatebox{45}{\textbf{GPT-4}} & \rotatebox{45}{\textbf{LLaMA 70B}} & \rotatebox{45}{\textbf{LLaMA 13B}} & \rotatebox{45}{\textbf{LLaMA 7B}} & \rotatebox{45}{\textbf{Mistral}} \\
\midrule
Baseline & 0.083560 & 0.0862 & 0.4289& 0.1401& 0.1284 & 0.0893 &0.395\\
Reason & 0.07552 & \textbf{0.000638} & 0.5739 &0.09302& 0.5313 & 0.2732  &0.1278\\
DEI & 0.6463 & 0.0527 & 0.1312 & 0.3834&0.1461 & 0.09356 & 0.9220\\
Reason + DEI & 0.05351 & 0.0789 & 0.1232 &0.3167 &0.1524 & 0.1124 & 0.5059\\
\bottomrule
\end{tabular*}

\vspace{1.5em}

\captionof{table}{Qualified Wilcoxon Test $P$-Value Results Between Male and Female for Pronouns Dataset}
\begin{tabular*}{\textwidth}{@{\extracolsep{\fill}}lccccccc}
\toprule
\textbf{Prompt} & \rotatebox{45}{\textbf{Gemini}} & \rotatebox{45}{\textbf{GPT-3.5}} & \rotatebox{45}{\textbf{GPT-4}} & \rotatebox{45}{\textbf{LLaMA 70B}} & \rotatebox{45}{\textbf{LLaMA 13B}} & \rotatebox{45}{\textbf{LLaMA 7B}} & \rotatebox{45}{\textbf{Mistral}} \\
\midrule
Baseline &\textbf{4.274e-03} & 0.5711 & \textbf{0.0373}&0.3057 & 0.3955 & 0.7205  &0.07539\\
Reason & 8.676e-03 & 0.3144 & \textbf{0.0461} &\textbf{0.029473}& \textbf{0.0254} & 0.4081 &\textbf{0.07701}\\
DEI & \textbf{8.883e-04} & 0.2422 & \textbf{0.003916} &\textbf{0.004424}& \textbf{0.000172} & 0.2195 & 0.1154\\
Reason + DEI & \textbf{8.294e-05} &\textbf{0.01160} & \textbf{0.001988} &\textbf{0.02890} &\textbf{0.000091} & 0.2283&  0.08218\\
\bottomrule
\end{tabular*}

\vspace{1.5em}

\captionof{table}{Qualified Wilcoxon Test $P$-Value Results Between Males and Non-binary for Pronouns Dataset}
\begin{tabular*}{\textwidth}{@{\extracolsep{\fill}}lccccccc}
\toprule
\textbf{Prompt} & \rotatebox{45}{\textbf{Gemini}} & \rotatebox{45}{\textbf{GPT-3.5}} & \rotatebox{45}{\textbf{GPT-4}} & \rotatebox{45}{\textbf{LLaMA 70B}} & \rotatebox{45}{\textbf{LLaMA 13B}} & \rotatebox{45}{\textbf{LLaMA 7B}} & \rotatebox{45}{\textbf{Mistral}} \\
\midrule
Baseline &\textbf{4.441e-06} & \textbf{0.02319} & \textbf{0.003925} &0.6518 &\textbf{0.01779}  &\textbf{0.03968} &\textbf{0.008235}\\
Reason & \textbf{9.769e-06} & \textbf{0.01559} & \textbf{0.01021} & 0.6171&\textbf{0.004211} & 0.0545 & 0.8159\\
DEI & \textbf{1.277e-04} & \textbf{0.001958} & \textbf{0.000010} &\textbf{0.04822} &\textbf{0.02128} & \textbf{0.003681} & 0.1389\\
Reason + DEI & \textbf{2.735e-09}  &\textbf{0.000021}  &\textbf{0.000004} &\textbf{0.2368} &\textbf{0.01307} & \textbf{0.005253}  &0.2820\\
\bottomrule
\end{tabular*}

\vspace{1.5em}

\captionof{table}{Qualified Wilcoxon Test $P$-Value Results Between Males and Females for Gendered Name Dataset}
\begin{tabular*}{\textwidth}{@{\extracolsep{\fill}}lcccccc}
\toprule
\textbf{Prompt} & \rotatebox{45}{\textbf{Gemini}} & \rotatebox{45}{\textbf{GPT-3.5}} & \rotatebox{45}{\textbf{LLaMA 70B}} & \rotatebox{45}{\textbf{LLaMA 13B}} & \rotatebox{45}{\textbf{LLaMA 7B}} & \rotatebox{45}{\textbf{Mistral}} \\
\midrule
Baseline & \textbf{0.01704} & 0.9380  &  \textbf{0.06947}&\textbf{0.07366} & \textbf{0.005192}  &0.977\\
Reason & \textbf{2.555e-03} & 0.9641    &  0.05957&\textbf{0.04660} & 0.08927  &0.2322\\
DEI &  \textbf{8.993e-04} & 0.2176      &  \textbf{0.01552}&\textbf{0.0189} & \textbf{0.004938} & 0.4036\\
Reason + DEI & 0.05642& 0.5066  &  \textbf{0.03334}&\textbf{0.001499} & 0.1836  &\textbf{0.004440}\\
\bottomrule
\end{tabular*}

\vspace{1.5em}

\captionof{table}{Compensation Kolmogorov-Smirnov Test $P$-Value Results for Pronouns Dataset (Gender: F \& NB)}
\begin{tabular*}{\textwidth}{@{\extracolsep{\fill}}lccccccc}
\toprule
\textbf{Prompt} & \rotatebox{45}{\textbf{Gemini}} & \rotatebox{45}{\textbf{GPT-3.5}} & \rotatebox{45}{\textbf{GPT-4}} & \rotatebox{45}{\textbf{LLaMA 70B}} & \rotatebox{45}{\textbf{LLaMA 13B}} & \rotatebox{45}{\textbf{LLaMA 7B}} & \rotatebox{45}{\textbf{Mistral}} \\
\midrule
Baseline & 0.07417 & \textbf{1.081e-06}  &0.651077 &\textbf{0.01781}& \textbf{0.000032} & \textbf{0.01645} & 0.08141\\
Reason &0.950872 & \textbf{1.723e-05}  &0.737398 &\textbf{2.283e-04}& \textbf{5.806e-05} & \textbf{9.986e-04} & 0.672845\\
DEI & 0.8146 & \textbf{5.454e-06}  &0.2493 &0.8724& \textbf{0.02434} & \textbf{0.01290}&  0.8885\\
Reason + DEI & \textbf{0.03179} & \textbf{3.783e-11}  &\textbf{4.595e-03} &0.9525 &0.08140 & \textbf{2.812e-06} & \textbf{0.02434}\\
\bottomrule
\end{tabular*}

\vspace{1.5em}

\captionof{table}{Compensation Kolmogorov-Smirnov Test $P$-Value Results for Pronouns Dataset (Gender: M \& F)}
\begin{tabular*}{\textwidth}{@{\extracolsep{\fill}}lccccccc}
\toprule
\textbf{Prompt} & \rotatebox{45}{\textbf{Gemini}} & \rotatebox{45}{\textbf{GPT-3.5}} & \rotatebox{45}{\textbf{GPT-4}} & \rotatebox{45}{\textbf{LLaMA 70B}} & \rotatebox{45}{\textbf{LLaMA 13B}} & \rotatebox{45}{\textbf{LLaMA 7B}} & \rotatebox{45}{\textbf{Mistral}} \\
\midrule
Baseline & 0.1762 & \textbf{4.366e-04}  &0.160365 & \textbf{3.1436e-19}&\textbf{0.038105} & \textbf{4.333e-15} & \textbf{0.015181}\\
Reason &  0.348324 & \textbf{8.153e-04}  &0.694540 & \textbf{8.603e-18}&0.482021 & \textbf{8.949e-10} & \textbf{0.002645}\\
DEI &  0.5807 & \textbf{1.347e-03} & 0.3533 & \textbf{1.286e-12}&\textbf{0.0440} & \textbf{6.822e-18} &\textbf{0.000204}\\
Reason + DEI & 0.7327 & \textbf{3.182e-08}  &0.05823 &\textbf{1.113e-10}& 0.9176 & \textbf{4.837e-13} & 0.6511\\
\bottomrule
\end{tabular*}

\vspace{1.5em}

\captionof{table}{Compensation Kolmogorov-Smirnov Test $P$-Value Results for Pronouns Dataset (Gender: M \& NB)}
\begin{tabular*}{\textwidth}{@{\extracolsep{\fill}}lccccccc}
\toprule
\textbf{Prompt} & \rotatebox{45}{\textbf{Gemini}} & \rotatebox{45}{\textbf{GPT-3.5}} & \rotatebox{45}{\textbf{GPT-4}} & \rotatebox{45}{\textbf{LLaMA 70B}} & \rotatebox{45}{\textbf{LLaMA 13B}} & \rotatebox{45}{\textbf{LLaMA 7B}} & \rotatebox{45}{\textbf{Mistral}} \\
\midrule
Baseline & 0.6023 & 0.1513  &\textbf{4.096e-02} &\textbf{3.194e-10}&0.119031 & \textbf{3.643e-28}  &0.2365\\
Reason & 0.331997 & 0.6076 & 0.998997 & \textbf{1.520e-05}&\textbf{4.595e-03} & \textbf{5.848e-23} & \textbf{4.37e-04}\\
DEI & 0.9603 & 0.1190314  &0.4429 &\textbf{6.505e-14} &0.8724&  \textbf{8.997e-31} & \textbf{2.28e-04}\\
Reason + DEI & \textbf{2.734e-02} & 0.6511  &\textbf{6.000e-06}& \textbf{5.938e-12}& 0.3877 & \textbf{1.272e-34} & \textbf{1.989e-03}\\
\bottomrule
\end{tabular*}

\vspace{1.5em}

\captionof{table}{Compensation Kolmogorov-Smirnov Test $P$-Value Results for Males and Females for Gendered Names Dataset}
\begin{tabular*}{\textwidth}{@{\extracolsep{\fill}}lcccccc}
\toprule
\textbf{Prompt} & \rotatebox{45}{\textbf{Gemini}} & \rotatebox{45}{\textbf{GPT-3.5}} & \rotatebox{45}{\textbf{LLaMA 70B}} & \rotatebox{45}{\textbf{LLaMA 13B}} & \rotatebox{45}{\textbf{LLaMA 7B}} & \rotatebox{45}{\textbf{Mistral}} \\
\midrule
Baseline & 0.3009 & \textbf{6.235e-02} &\textbf{1.060e-09}& 0.1119   & \textbf{1.462e-04} & 0.2493\\
Reason & 0.6022 & \textbf{7.342e-05} &\textbf{2.575e-16} &0.3369 & \textbf{4.197e-03} & \textbf{0.03055}\\
DEI & 0.07417 & 0.9176   & \textbf{5.557e-09}&0.2365   & \textbf{3.832e-03} & 0.5227\\
Reason + DEI& \textbf{0.0426} & 0.4429 & \textbf{8.949e-10}&\textbf{0.002188} & \textbf{1.306e-04} & 0.2493\\
\bottomrule
\end{tabular*}

\end{onecolumn}
\twocolumn

\end{document}